\documentclass[useAMS,usenatbib]{mn2e}
\usepackage{graphicx,amssym,color}
\newcommand{\bc}{\begin{center}}
\newcommand{\ec}{\end{center}}

\newcommand{\kpc}            {\,{\rm kpc}}

\newcommand{\hkpc}           {\,h^{-1}\,{\rm kpc}} 

\newcommand{\hMsun}          {\,h^{-1}\,{\rm M}_\odot}
\newcommand{\kms}            {\,\,{\rm km}\,\,{\rm s}^{-1}}

\title[Assembly and Structure of Dark Matter Halos] { Assembly History
  and  Structure of Galactic Cold Dark Matter Halos} 
\author[Wang et al.]{\parbox{18cm}{
J.~Wang$^{1,2}$,
J.~F.~Navarro$^{3}$,
C.~S.~Frenk$^{1}$, 
S.~D.~M.~White$^{4}$, 
V.~Springel$^{4,5}$, 
A.~Jenkins$^{1}$, 
A.~Helmi$^{6}$ %
A.~Ludlow$^{7},$ %
and M.~Vogelsberger$^{8}$
}\vspace{0.3cm}\\
$^1$ Institute for Computational Cosmology, Dep. of Physics, Univ. of Durham, South Road, Durham  DH1 3LE, UK\\
$^{2}$ {Newton International Fellow}\\
$^{3}$ {Dep. of Physics \& Astron., University of
    Victoria, Victoria, BC, V8P 5C2, Canada}\\
$^{4}$ {Max-Planck-Institut f\"{u}r Astrophysik,
Karl-Schwarzschild-Stra\ss{}e 1, 85740 Garching bei M\"{u}nchen,
Germany}\\
$^{5}$ {Heidelberg Institute for Theoretical Studies, Schloss-
Wolfsbrunnenweg 35, 69118 Heidelberg, Germany}\\
$^{6}$ {Kapteyn Astronomical Institute, Univ. of Groningen,
P.O. Box 800, 9700 AV Groningen, The Netherlands}\\
$^{7}$ {Argelander-Institut f\"{u}r Astronomie, Auf dem H\"{u}gel 71,
  D-53121 Bonn, Germany}\\
$^{8}$ {Harvard-Smithsonian Center for Astrophysics, 60 Garden Street,
  Cambridge, MA, 02138, USA}\\
}

\begin{document}



\maketitle
\label{firstpage}
\begin{abstract}
  We use the Aquarius simulation series to study the imprint of
  assembly history on the structure of Galaxy-mass cold dark matter
  halos. Our results confirm earlier work regarding the influence of
  mergers on the mass density profile and the inside-out growth of
  halos. The inner regions that contain the visible galaxies are
  stable since early times and are significantly affected only by
  major mergers. Particles accreted diffusely or in minor mergers are
  found predominantly in the outskirts of halos. Our analysis 
  reveals trends that run counter to current perceptions of
  hierarchical halo assembly. For example, major mergers (i.e. those
  with progenitor mass ratios greater than 1:10) contribute little to
  the total mass growth of a halo, on average less than 20\% for our
  six Aquarius halos. The bulk is contributed roughly equally by minor
  mergers and by ``diffuse'' material which is not resolved into
  individual objects. This is consistent with modeling based on
  excursion-set theory which suggests that about half of this diffuse
  material should not be part of a halo of {\it any} scale.
  Interestingly, the simulations themselves suggest that a significant
  fraction is not truly diffuse, since it was ejected from earlier
  halos by mergers prior to their joining the main system. The
  Aquarius simulations resolve halos to much lower mass scales than
  are expected to retain gas or form stars. These results thus confirm
  that most of the baryons from which visible galaxies form are
  accreted diffusely, rather than through mergers, and they suggest
  that only relatively rare major mergers will affect galaxy structure
  at later times.
\end{abstract}


\section{Introduction}
\label{SecIntro}

Hierarchical growth is a signature prediction of the $\Lambda$CDM cosmogony,
our current standard picture of cosmic structure formation. $\Lambda$CDM
postulates a flat universe with a cosmological constant, cold dark
matter, and gaussian initial conditions generated at very early
times. The basic units of nonlinear structure are dark matter halos
that grow by accretion and merging as gas cools and condenses into
galaxies in their cores. The statistics of this process are amenable
to analytic modeling, which can in turn be validated and extended
through cosmological N-body techniques \citep[see,
e.g.,][]{Press1974,bond91,kauffmann93,lacey93,Cole1996,Efstathiou1988,Jenkins2001}.

The mass function of collapsed structure and its evolution with time, 
the clustering of halos of different mass, the origin and character of
scaling laws relating halo properties are all results that can be
understood within the context of the excursion-set modeling making
reference only to the initial power spectrum of density fluctuations
and to the universal expansion history \citep[see, e.g.,][]{lacey93,Mo1996,NFW96}.

The $\Lambda$CDM power spectrum, $P(k)$, can be computed in detail using
linear theory.  Under the simplifying assumption that the
``temperature'' of the dark matter is zero (or, equivalenty, that the
dark matter particle, if a thermal relic, has ``infinite''
mass), $P(k)$ approaches $k^{-3}$ on the smallest scales with a mass
variance that diverges logarithmically there. Excursion-set theory
then predicts that, at times of interest, effectively all of the mass
of the Universe is in clumps of some mass. The assembly of a halo thus
consists, in this simplified case, merely of the merging of the
myriads of smaller mass subhalos that collapsed at earlier times.

Under these conditions merging is the basic engine of halo growth.
The rapid mixing driven by the violently fluctuating potential of a
merger has awesome transformative powers. Mergers can erase, at least
partially, memory of the initial conditions and leave remnants whose
broad structure is roughly independent of the cosmological conditions
of formation \citep{White1978,vanAlbada1982}. 
Not all mergers, however, are created equal, and it has long been
appreciated that the effects of major mergers differ qualitatively
from those of minor events. In particular, major and minor mergers
affect differently the internal structure of the remnant. Major
mergers lead to ``rapid growth'' in the mass of an object, which has
been linked with radical changes in the halo structural
parameters. Minor mergers, on the other hand, are associated with
``slow growth'' evolutionary phases that leave the inner structure of
the main halo relatively intact and affect mainly the periphery of the
remnant \citep{Salvador1998,wechsler02, zhao03a,Tasitsiomi2004,Diemand2007}.

The idea of merging as the exclusive mechanism of halo growth has received
some backing in the literature, most recently from \citet{Madau2008},
who report that most of the mass in their Via Lactea simulation of a
galaxy-sized halo ``is acquired in resolved discrete clumps, with no
evidence for significant smooth infall''. On the other hand, a number of
recent papers have also argued that the fraction of mass accreted
``diffusely'' might be substantial \citep[see, e.g.,][and references
therein]{Fakhouri2010,Angulo2010,Genel2010}.

\citet{Angulo2010}, in particular, note that when a realistic cold dark
matter particle candidate is chosen its small but non-negligible thermal
velocity introduces a cutoff (and finite variance) in the power
spectrum on small scales that can have a profound impact on the way
the evolving hierarchy of collapsed structures develops. Working
through the numbers appropriate for a neutralino-dominated Universe,
these authors argue that, as late as $z\sim 20$, most of the mass of
the Universe is not yet part of {\it any} halo. These authors also
argue that a typical galaxy-sized halo accretes at least $10\%$ of its
mass in diffuse form.

If these numbers are correct the actual fraction of smoothly-accreted
material in a typical N-body halo must be much higher, since
simulations can only resolve a limited range of nonlinear scales and
a fair fraction of the mass is expected to be locked up in unresolved
small mass clumps. \citet{Angulo2010} argue that, even in the best
simulations currently available, up to $30$-$40\%$ of the mass of a
galactic halo could have been accreted in diffuse form, in clear
disagreement with the results of \citet{Madau2008}. It is clearly
important to resolve this disagreement, especially given the
importance of diffuse mass accretion for galaxy formation emphasized
in recent papers \citep[see, e.g.,][and references
therein]{Keres2005,Dekel2009}.

We address these issues here using the N-body simulations of the
Aquarius Project \citep{Springel2008a}. This simulation series follows
the formation of six different $\Lambda$CDM halos at various resolutions, and
includes the best-resolved galactic dark matter halo simulated so far,
an object with more than one billion particles within the virial
radius. We begin with a brief description of the Aquarius Project in
Sec.~\ref{SecNumSims}, and move on in Sec.~\ref{SecTom} to a
systematic study of the radial structure of halos in terms of the mass
of their progenitor halos and the time of their
accretion/merging. Sec.~\ref{SecAcc} considers the mode of accretion
into the halos in detail and addresses the fraction of mass accreted
in diffuse form. We end with a brief summary of our main conclusions
in Sec.~\ref{SecConcl}.

\begin{figure}
\bc
\hspace{-1.cm}
\resizebox{9cm}{!}{\includegraphics{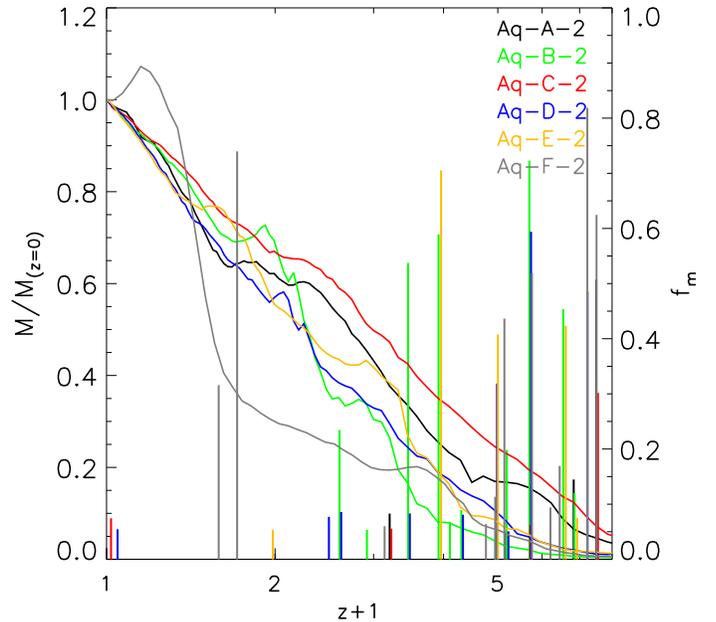}}\\%
\caption{The evolution of the mass of the main friends-of-friends
  (FOF) progenitor of the six level-2 Aquarius halos. The curves show
  the mass in units of the mass at $z=0$ (labels on the left
  $y$-axis). Vertical segments indicate the mass ratio of the largest
  merger event occurring at each snapshot (labels on the right
  $y$-axis). Only merger events with mass ratio exceeding $0.05$ are
  shown. Colors identify individual halos, as labelled in the
  figure. Note that only halo Aq-F-2 has undergone a major ($f_m>0.1$)
  merger after $z=1$.}
\label{fig:mvsz}
\ec
\end{figure}

\begin{figure}
\bc
\hspace{-1.cm}
\resizebox{9cm}{!}
{\includegraphics{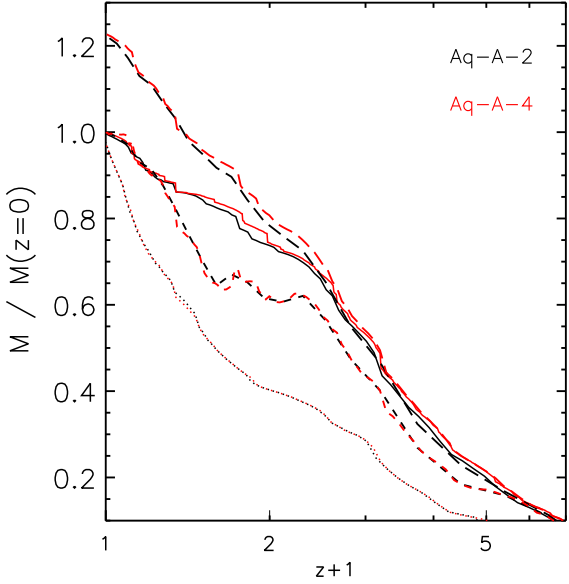}}\\%
\caption{Mass accretion history of halo Aq-A. Curves in black
  correspond to the level-2 resolution halo, in red to the level-4
  run. All masses are normalized to the FOF mass of the halo at
  $z=0$. Four definitions of halo mass buildup are compared. The
  long-dashed (top) curves indicate the mass of all particles
  ``associated'' with the halo; i.e., those that were part of the main
  progenitor at {\it any time} before redshift $z$. The dotted
  (bottom) curves are the mass of particles that belong to the most
  massive progenitor {\it at all times} after $z$ (i.e., after the
  time of last accretion). The thin solid curves (second set from top)
  indicate the cumulative mass of particles as a function of the
  redshift of {\it first} accretion into the main progenitor
  (regardless of whether they leave and re-enter the main progenitor
  subsequently). The short-dashed curves indicate the conventional FOF
  mass of the main progenitor at each time. This comparison illustrates the fact
  that a significant fraction of the mass of the main progenitor is
  pushed out of the halo boundary during its evolution; much of it is
  re-accreted later, but more than $20\%$ is still outside the main
  halo at $z=0$.}
\label{fig:acccomp}
\ec
\end{figure}

\section{The Numerical Simulations}
\label{SecNumSims}

The Aquarius Project \citep{Springel2008a} consists of a suite of
large N-body simulations of six dark matter halos of mass consistent
with that expected for the halo of the Milky Way.  Our simulations
assume the $\Lambda$CDM cosmology, with parameters consistent with the {\small
  WMAP} 1-year data \citep{spergel03}: matter density
parameter, $\Omega_{\rm M}=0.25$; cosmological constant term,
$\Omega_{\Lambda}=0.75$; power spectrum normalisation, $\sigma_8=0.9$;
spectral slope, $n_s=1$; and Hubble parameter, $h=0.73$.

The halos were identified in a $900^3$-particle N-body simulation of a
cubic volume $100\, h^{-1}$ Mpc on a side, a lower resolution version
of the Millennium-II Simulation \citep{Boylan-Kolchin2009}. This
volume was resimulated using exactly the same power spectrum and
phases of the original simulation, but with additional high-frequency
waves added to regions surrounding the initial Lagrangian volume of
each halo. The high-resolution region was populated with low-mass
particles and the rest of the volume with particles of higher
mass. These ``zoomed-in'' simulations of selected regions or
individual objects have become common practice to make galaxies; for
details we refer the reader to \citet{Power2003}.

The six Aquarius halos are labelled ``Aq-A'' through ``Aq-F''. Each
was resimulated at different resolutions in order to assess numerical
convergence. A suffix, $1$ to $5$, identifies the resolution level,
with level 1 denoting the highest resolution. Between levels 1 and 5,
the particle mass ranges from $m_p \sim 2 \times 10^3 \, M_\odot$ to $\sim
3 \times 10^6 M_\odot$. Level 1 was performed only for Aq-A and
contains roughly $1.1$ billion particles within the virial radius. All
six halos were simulated at level-2 resolution. Each of these has more
than $100$ million particles within the virial radius.

In this study, we analyze primarily the level-2 simulations but we
also use lower resolution versions of Aq-A to test for numerical
convergence. For the level-2 simulations, the particle mass is
$m_{p}\simeq 1 \times 10^4\hMsun$ and the softening length is $
\epsilon= 48 h^{-1} {\rm pc}$. At $z=0$, the six haloes have a similar ``virial''
mass, $M_{200}\sim 1$-$2 \times 10^{12}\hMsun$, where $M_{200}$ is the
mass contained within $r_{200}$, the radius of a sphere of mean
density 200 times the critical density for closure{\footnote{This
    choice defines implicitly the virial radius of the halo,
    $r_{200}$, and its virial velocity, $V_{200}$.}}. The circular
velocity curve of the halos peaks at roughly $V_{\rm max} =220 \pm40
\kms$.  For further details of the Aquarius Project, we refer the
reader to \citet{Springel2008b} and \citet{Navarro2010}.

At every snapshot in the simulation we find nonlinear structures
using the friends-of-friends (FOF) algorithm of \citet{davis85}, with
a linking length of 0.2 times the mean interparticle separation and
$32$ particles as the minimum number of particles per group. We then
construct a merger tree for the final Aquarius halos linking FOF
progenitors at each time. We also identify bound substructures within
each FOF halo (subhalos) using the {\small SUBFIND} algorithm
of \citet{Springel2005}. Merger trees for subhalos are constructed as
described in \citet{Springel2008b}.

For simplicity, unless otherwise explicitly noted, we shall identify a
halo with the FOF structure that contains it. Note that the mass of
FOF halos does not necessarily coincide with the virial mass alluded
to above (FOF structures are larger; they typically enclose a halo and
a small part of its surroundings); we shall comment on these
differences when appropriate in the analysis that follows.

Fig.~\ref{fig:mvsz} shows the growth of the FOF mass of the main
progenitor of each halo, normalized to its value at the present time,
$z=0$. Merger events with mass ratio greater than $f_m=M_{\rm
  prog}/M_{\rm main}=0.05$ are noted by vertical lines in the colour
corresponding to the appropriate halo (scale on the right).

If we define the formation time of a halo as the time when the main
progenitor first reaches half the final mass, then the formation time
of the halos is $z\sim 1.2 - 2.2$, except for halo Aq-F-2 which is
clearly different from the rest. Its formation redshift is $z=0.6$,
when its mass almost doubles as a result of an almost equal-mass
merger (the actual mass ratio of the two progenitors is
$f_m=0.75$). The other five halos have similar mass growth histories
but different merger histories. For example, halos B, D, and E
experienced major mergers at high redshift ($z\sim 2 - 7$), while
halos A and C grew in a relatively quiescent fashion and did not
experience any major mergers after $z=6$. Although the Aquarius
haloes have similar final masses, they have varied formation
histories, which 
\cite{boylan10} has shown, sample the range of behaviours seen in the
Millennium~II simulation for halos of this mass. The Aquarius haloes
therefore provide a suitable sample to study the diversity in assembly 
histories of halos similar to that surrounding the Milky Way.

\begin{figure*}
\bc
\hspace{-1.4cm}
\resizebox{21cm}{!}{\includegraphics{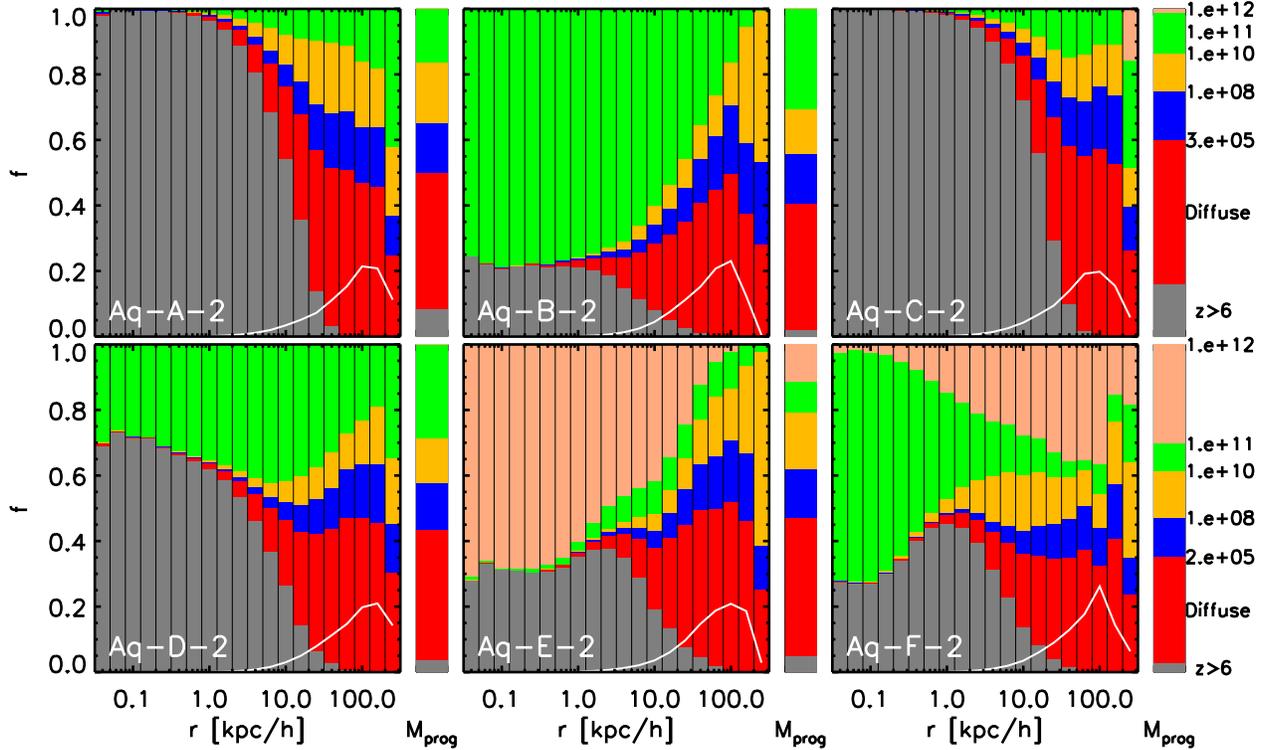}}\\%
\caption{The radial distribution of particles in the FOF $z=0$ halo 
colour-coded 
  according to $M_{\rm prog}$, the mass of the progenitor to which
  each particle belonged at the time of (first) accretion into the
  main halo.  The bars represent the fraction of the mass in each
  spherical shell brought in by halos with mass in the range indicated
  by the key to the right of each panel. This key also gives the total
  fraction (summed over all radial shells) of mass brought in by
  different progenitors. Material accreted before $z=6$ is indicated
  in grey.  Diffuse material, that is, particles that were not part of
  any FOF halo at the time of accretion, are indicated in red.  Masses
  are in units of $\hMsun$. The white curve gives the fraction of the
  total FOF halo mass in each radial shell.}
\label{fig:progmassprof} 
\ec
\end{figure*}

\begin{figure*}
\bc
\hspace{-1.4cm}
\resizebox{21cm}{!}{\includegraphics{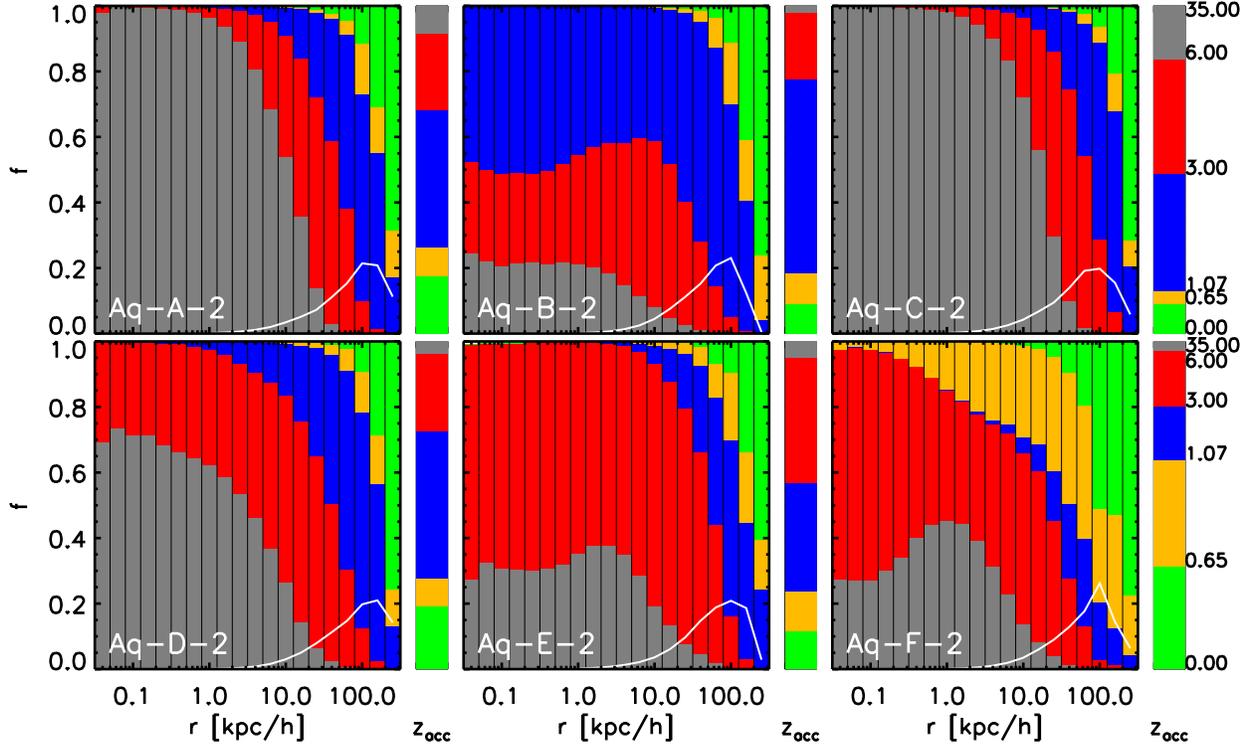}}\\
\caption{As Fig~\ref{fig:progmassprof}, but for $z_{\rm acc}$, the
  redshift of (first) accretion of particles into the main halo. Different
  colours indicate accretion redshift in the intervals indicated by
  the key to the right of each panel. This key also gives the total
  mass fraction (summed over all radial shells) accreted in each
  redshift interval.  The white curve gives the fraction of the total
  FOF halo mass in each radial shell.}
\label{fig:taccprof}
\ec
\end{figure*}

\section{Mass and Accretion History}
\label{SecTom}

\subsection{Mass growth and definition of accretion}
\label{SecAccDef}

Starting at $z=0$ we identify the main trunk of the merger tree of
each of our halos by stepping back in time, defining the main
progenitor at time $n-1$ to be the largest FOF halo which is a
progenitor of the main progenitor at time $n$.  For each particle in
the final object at $z=0$, we register the redshift when it was
accreted into the main progenitor (i.e. when it first ceased to be a
single particle or a member of {\it another} FOF group), $z_{\rm
  acc}$, and the mass of the FOF group, $M_{\rm prog}$, to which it
belonged at the snapshot immediately preceding the time of
accretion. For ``diffuse'' or ``smooth'' accretion, terms we use
interchangeably throughout, $M_{\rm prog}$ equals the mass of a single
particle.

This seemingly straightforward definition of accretion is complicated
by the fact that some particles can leave the main progenitor and be
re-accreted again later on. This process can actually recur multiple
times, and is usually associated with accretion events, where a small but
non-negligible fraction of the mass is propelled into highly energetic
orbits \citep{Balogh2000,Gill2005,Diemand2007,Ludlow2009}.

There is therefore some ambiguity in the meaning of accretion time whose
effects we illustrate in Fig.~\ref{fig:acccomp}.  Here we compare the
growth of halo Aq-A using several plausible definitions of accretion and
two different levels of resolution to check for possible numerical
artifacts. The dashed curves (second from the bottom) track, as in
Fig.~\ref{fig:mvsz}, the conventional FOF mass of the main progenitor,
normalized to its value at $z=0$. The long-dashed curves (top) show,
on the other hand, the cumulative mass of all particles that, at any
time before $z$, have been part of the main progenitor. These
``associated'' particles exceed the FOF halo at $z=0$ by more than
$20\%$, highlighting the importance of the energy redistribution
process described in the previous paragraph.

The solid thin curves, on the other hand, track the $z=0$ FOF
particles, but use the time of {\it first} accretion to define
$z$. The difference between this and the conventional FOF mass is a
direct indicator of the accretion-escape-reaccretion process alluded
to above. Finally, the dotted (bottom) curves use the time of {\it
  last} accretion of particles in the $z=0$ FOF
group. Fig.~\ref{fig:acccomp} shows clearly that a halo is a dynamic
object and not a static ``bucket'' of mass that gets progressively
filled by accretion. For simplicity we shall in what follows adopt the
time of first entry as our default definition of accretion but we
caution that other definitions of accretion may on occasion be more
useful, depending on the aim of the analysis. The good agreement
between the results for Aq-A-4 and Aq-A-2 show that these conclusions
are insensitive to the numerical resolution of the simulations.


\begin{figure}
\bc
\hspace{-1.4cm}
\resizebox{9cm}{!}
{\includegraphics{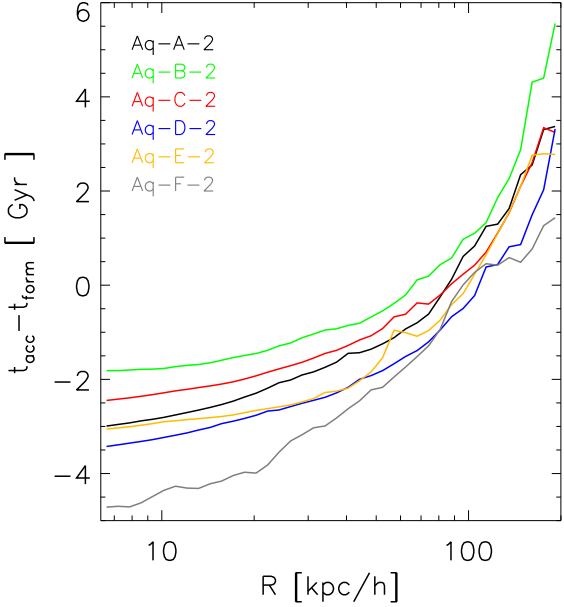}}\\%
\caption{The mean accretion time of particles in different radial
  shells at $z=0$. Accretion times are shown relative to the formation
  time of each halo, $t_{\rm form}$, defined as the time when the halo
  first reached half its final mass. When defined this way the mean
  accretion time profile is similar for all Aquarius halos.}
\label{fig:rvstacc} 
\ec
\end{figure}

\begin{figure}
\bc
\hspace{0.2cm}
\resizebox{9cm}{!}
{\includegraphics{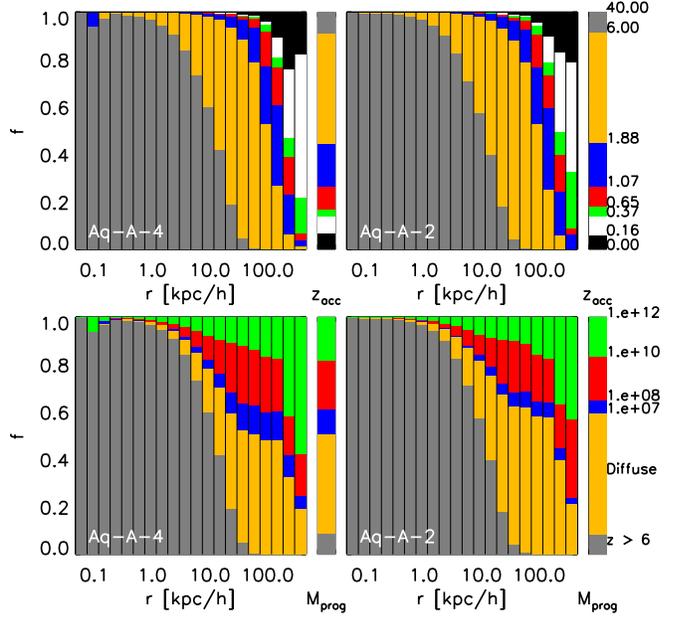}}\\%
\caption{A convergence test of the radial gradients in progenitor mass
  and accretion redshift. The top two panels are analogous to
  Fig.~\ref{fig:taccprof}, the bottom panels to
  Fig.~\ref{fig:progmassprof}, but for the level-2 and level-4 Aq-A
  runs. These runs differ solely in numerical resolution; level-2
  runs have $30\times$ more particles and $5\times$ smaller
  gravitational softening than their level-4 counterparts. The
  excellent agreement shows that the results
  presented in Figs.~\ref{fig:progmassprof} and ~\ref{fig:taccprof}
  are insensitive to numerical resolution.}
\label{fig:convprof} 
\ec
\end{figure}

\begin{figure}
\bc
\hspace{-1.4cm}
\resizebox{9cm}{!}
{\includegraphics{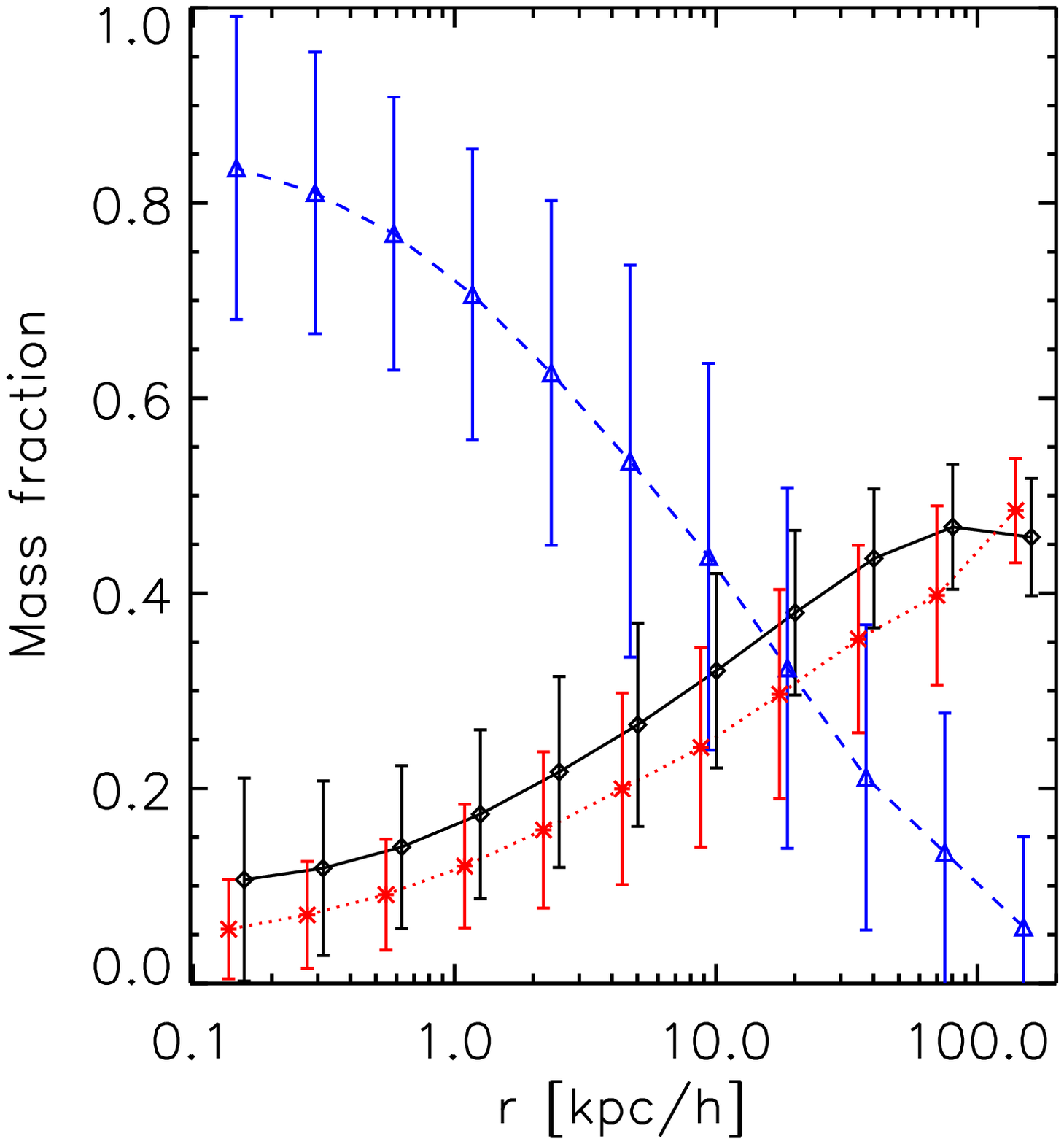}}\\%
\caption{The fraction of particles in a series of spherical shells
  that were accreted smoothly (black circles -- solid line), by minor
  mergers (starred symbols -- dotted) and by major mergers (open
  triangles -- dashed). The distinction between minor and major mergers
  is made at a mass ratio of $10$:$1$. Fractions are averaged over all
  six level-2 Aquarius halos and error bars show the {\em rms} scatter among
  halos.  Despite substantial scatter, the trends are clear. Major
  mergers contribute to the inner regions, diffuse accretion and minor
  mergers mainly to the outer regions which contain the bulk of the
  mass.}
\label{fig:mrgdiffacc}
\ec
\end{figure}


\subsection{Progenitor mass distribution as a function of radius}
\label{sec:mahma}

We now investigate the fate of particles accreted in progenitor halos
of different mass. More specifically, we investigate the distribution
of halo mass in spherical shells (centred at the location of the
minimum in the gravitational potential) and apportion the contribution
according to the mass, $M_{\rm prog}$, of the subhalos that brought
each particle into the main progenitor.

This is shown in Fig.~\ref{fig:progmassprof}, which gives, in each
panel, the fraction of particles in concentric radial shells of each
$z=0$ Aquarius halo, split into six mass bins, according to the mass
of the progenitor halo at the time of accretion.  Each bin is
identified by a different colour according to the key to the right of
each panel. This key, in addition, gives the total fraction (summed
over all radial shells) of mass brought in by progenitors of different
mass. Material that was already in place at $z=6$ (when few resolved
progenitor halos exist) is indicated in grey; in this section and the
next we do not consider the assembly history at earlier times.
Particles that in the snapshot preceding accretion are unattached to
any resolved halo (i.e., diffuse accretion) are indicated in red.

Fig.~\ref{fig:progmassprof} shows that there is considerable
halo-to-halo variation in the mass spectrum of the progenitors of the
final halo.  Consider, for example, the integrated halo mass. As the
key on the right of each panel shows, the fraction of mass that was
already in place at $z=6$ ranges from $\sim 5$ to $\sim 15\%$.  The
fraction of mass accreted diffusely is substantial in all cases,
ranging from $\sim 30$ to $\sim 40\%$,  The diffuse fraction increases 
with radius, from a few per cent within $10 \hkpc$, to more than 20
per cent within $100\hkpc$. This agrees with the results found by 
\citet{helmi02} from an N-body simulation of a cluster halo scaled to 
a galactic mass. The radial behaviour is diverse. For example, in
halos A and C the material in the innermost 
region, $r < 1\kpc$, was already in place before $z=6$ and has
undergone little change since. By contrast, in halos B, E and F,
most of the central mass was brought in after $z=6$ through
mergers involving host halos with mass greater than
$10^{10}\hMsun$. 

Halo D is intermediate between these two extremes. Large progenitors,
of mass $>10^{10}\hMsun$, bring in between $20$ and $40\%$ of the
final mass at all radii.  In general, the larger the mass of the carrier halo, the
greater the probability of the particles ending up in the central
regions. Very little of the diffuse material makes it into the central
regions, $r<10\hkpc$; indeed, most of it stays in the outer regions
where it typically contributes most of the mass.

\subsection{Accretion time distribution as a function of radius}
\label{sec:mahti}

We now turn our attention to the radial distribution of particles as a
function of their accretion time.  Analogously to
Fig.~\ref{fig:progmassprof}, Fig.~\ref{fig:taccprof} shows the
distribution of accretion redshifts, binned according to distance to
the centre of the halo. In each radial shell, the contribution from
material accreted in different time intervals is shown by strips coloured
according to the key shown to the right of each panel. The key also
gives the fractions of the total halo mass (summed over all shells)
accreted in each redshift interval.

The inside-out nature of halo assembly is clearly apparent in
Fig.~\ref{fig:taccprof}. On average, the peak contribution from each
accretion redshift interval marches outwards with time. The inner
regions are populated mostly by particles that were accreted early;
the outer layers were added gradually later. The cores of halo A and C
were in place before $z=6$ and evolved little thereafter. In halos D,
E, and F, the core particles were accreted by $z=3$, but for halo B,
the core is accreted at $z=1$ because two major mergers in the
redshift interval $1<z<3 $ bring in almost $50\%$ of the core mass
(see Fig.~\ref{fig:mvsz}).  These mergers happen relatively early,
while the total halo mass is small, and disrupt the original core
which then reforms from the new material. By contrast, the late major
merger undergone by halo $F$ has a relatively minor effect on the
core, probably because of the orbital parameters of the merger. The
core in this case is actually made primarily out of material that was accreted
in earlier, lesser mergers at $z\sim 4$.

The radial dependence of accretion time is quantified further in
Fig.~\ref{fig:rvstacc}, which shows the average accretion time as a
function of radius. Accretion time is plotted relative to the
formation time of the halo, $t_{\rm form}$, defined as the time when
the main progenitor first reaches half its final mass. With this
normalization, the radial dependence of the accretion time is
fairly similar for all Aquarius halos. On average, the material
in the inner $10\hkpc$ is assembled $2$ to $4$ Gyr before $t_{\rm form}$
whereas the material beyond $100\hkpc$ falls in $2$ to $4$ Gyr {\it after}
$t_{\rm form}$.  This onion-like growth is generic for cold dark
matter halos of galactic scale; it was seen also in a scaled cluster 
N-body simulation by \citet{helmi03}.

\subsection{Numerical convergence}
\label{sec:mahte}

Before discussing these results further we should verify that the
trends presented above are not unduly influenced by numerical
resolution. The availability of simulations of the same halo at varying
resolution allows for direct testing of the reliability of our
results. We do this by comparing the level-2 simulation of halo Aq-A,
which is the one we have analyzed so far, with its level-4
counterpart. The level-4 simulation has about $30\times$ poorer mass
resolution and $5\times$ poorer spatial resolution (softening).

The test is carried out in Fig.~\ref{fig:convprof}, which shows the
distributions of accretion time and the mass spectrum of progenitor
halos as a function of radius. These figures are analogous to
Figs.~\ref{fig:progmassprof} and~\ref{fig:taccprof}. Panels on the
left show the level-4 results, those on the right the level-2 results.
It is clear that the convergence of these properties is
excellent. There is no discernible difference in the distributions of
$z_{\rm acc}$ and at most a $\sim 10\%$ difference in the
distributions of $M_{\rm prog}$ in the lowest mass range, $M_{\rm prog} <
10^{7} \hMsun$.

\section{Modes of Accretion}
\label{SecAcc}

In this section we study how the growth of halos is apportioned
between major mergers, minor mergers, and smooth accretion; how the
material added in these modes is distributed in radius in the final
halos; and how much variation there is between halos. We will adopt a
FOF mass ratio of $10$:$1$ as our standard division between major and
minor mergers, although we will also give some results for the
stricter $3$:$1$ ratio adopted as a boundary by some authors. The fact
that we limit our FOF group catalogues to systems with at least 32
particles means that the boundary between minor mergers and ``smooth''
accretion occurs at a mass ratio of about $10^{6.5}$:$1$ at $z=0$
dropping to about $10^{5.5}$:1 at $z=4$ and to even smaller values at
higher redshifts.  In this section we consider increases in mass
through each of these growth modes throughout the entire history of
each halo, rather than halting at $z=6$ as in previous sections.

\subsection{Major mergers vs minor mergers}
\label{SecMajMinMrg}

In Fig.~\ref{fig:mrgdiffacc} we illustrate how major mergers, minor
mergers and diffuse accretion contribute to the $z=0$ mass in a series
of spherical shells, each spanning a factor of two in radius. The
symbols joined by lines give results averaged across the six Aquarius
halos, while the error bars indicate the {\it rms} scatter among
halos. Within $\sim 10\, h^{-1}$~kpc, major mergers are the dominant
source of the material, providing typically 40\% of the mass, while
minor mergers and smooth accretion bring in about 30\% each
respectively; within $\sim 1.0\, h^{-1}$~kpc, major mergers contribute
more than two thirds of the mass. Note, however, that less than $10\%$
of halo mass lies within $\sim 10\, h^{-1}$~kpc and less than $1\%$
lies within $\sim 1.0\, h^{-1}$~kpc. Note also from
Fig.~\ref{fig:taccprof} that the great majority of these major mergers
occurred at $z>3$ and many of them at $z>6$. Only in halos B and F are
there substantial contributions to these regions from major mergers at
redshifts below 3. The large error bars on these points indicate that
the scatter of the major merger contribution to the inner regions of
halos is large.

Beyond $10\, h^{-1}$ kpc, in the region which contains the bulk of the halo mass,
both minor mergers and diffuse accretion contribute more to halo growth than
major mergers. Indeed, averaged over all six halos, major mergers contribute
only $17\%$ of the total mass growth, with the values for individual halos
ranging from $3\%$ (Aq-A) to $36\%$ (Aq-F). For a stricter definition of a
major merger, requiring a mass ratio of $3$:$1$ or less, the mean major
merger contribution drops to just $9\%$, with individual values ranging from
$<0.1\%$ (Aq-A, Aq-C) to $25\%$ (Aq-F). Thus, major mergers are typically a small
contribution to overall halo growth.  The rest is split almost evenly between
minor mergers and ``diffuse'' accretion. It is interesting that the scatter in 
each of these contributions is very close to half that in the major merger
contribution. This shows that the minor merger and diffuse fractions fluctuate
up and down together, with minor mergers contributing slightly less than
half of the material not accounted for by major mergers at each radius and in
each halo.

\subsection{Diffuse accretion}
\label{SecDiffAcc} 

Given the conflicting claims in the literature regarding the
importance of diffuse accretion discussed in the Introduction, it is
important to explore possible biases and subtleties involved in
reckoning the amount of diffuse mass accreted. The dynamic nature of
halo buildup highlighted above (Sec.~\ref{SecAccDef}) introduces
amibiguities in the meaning of accretion, so we compare four
alternative definitions of $f_{\rm smooth}$, the total fraction of
mass in the FOF halo at $z=0$ that has been added smoothly:

I: all particles that were not part of {\it any} $32+$ particle FOF
group {\it in the snapshot immediately before} the time of first
accretion, $z_{\rm acc}$;

II: all particles that were never part of {\it any} $32+$ particle
{\it bound} structure (as identified by {\small SUBFIND}) before
$z_{\rm acc}$;

III. same as (II) but for $20+$ particles;

IV: same as (I) but for {\it all} snapshots before $z_{\rm acc}$.

\begin{figure}
\bc
\hspace{-1.cm}
\resizebox{9cm}{!}{\includegraphics{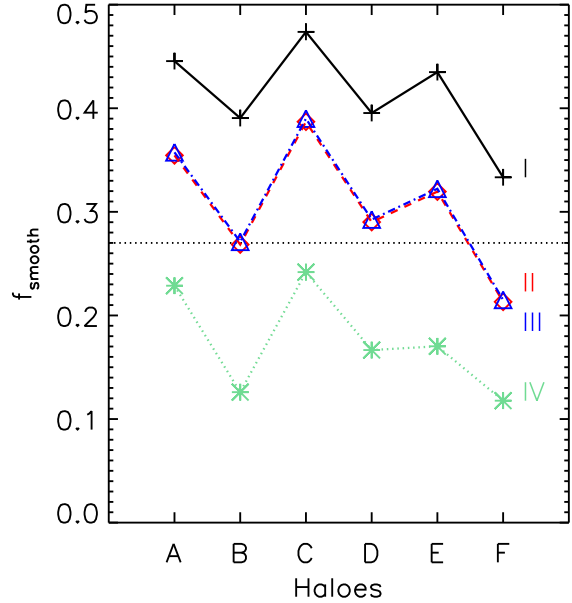}}\\%
\caption{The fraction of mass accreted in diffuse form in each level-2
  Aquarius halo. The $x$-axis lists the name of each halo. Each curve
  corresponds to one of the definitions of ``diffuse accretion''
  introduced in the text. Briefly, (I) are particles that are
  unattached to any FOF group identified in the snapshot immediately
  before first accretion; (II) refers to particles that do not belong
  to any {\it bound} structure with $\ge 32$ members in {\it any}
  snapshot before first accretion; (III) is as (II) but for $20$
  members; and (IV) denotes material that did not belong to {\it any}
  $N\ge 32$ FOF group at any time before first accretion. All curves
  use a total of 128 snapshots to estimate $f_{\rm smooth}$. The
  excursion-set prediction for a halo of the same mass and comparable
  numerical resolution is shown by the horizontal dotted line.}
\label{fig:diffacc}
\ec
\end{figure}

Fig.~\ref{fig:diffacc} compares results for the six level-2 Aquarius
halos. Criterion~(I), probably the simplest, is seen to give the
largest estimate of $f_{\rm smooth}$ in all cases. This criterion omits
those particles that were part of FOF halos in the past, but that have
left them and are unattached to any resolved structure just before
accretion. These make a surprisingly large fraction (about {\it
half}~!) of the smooth fraction computed using criterion~(I), as shown
by the bottom curve corresponding to criterion~(IV).

One shortcoming of criterion (IV), however, is the possibility that
FOF groups may artificially link in physically unrelated
particles. This is especially true in small-N groups \citep[for a
recent discussion, see, e.g.,][]{Bett2007}. Criteria (II) and (III)
account for this by requiring particles to be part of {\it bound}
structures; varying the threshold from $20$ to $32$ particles has
negligible effect on the results. This extra condition is seen to
increase $f_{\rm smooth}$ by roughly $50\%$ relative to criterion
(IV).  The contribution of diffuse accretion seems, therefore, to be
genuinely high, between $20$ and $40\%$ of the final halo mass
overall.

We note that, strictly speaking, $f_{\rm smooth}$ depends on the total
number of snapshots used in its estimation. The numbers quoted above
are based on a total of $128$ snapshots, but for one of the runs
(Aq-A-2) data were stored for $1024$ snapshots.  The estimate of
$f_{\rm smooth}$ according to criterion~(I) changes little when
considering $1024$ or $128$ snapshots: from $41\%$ to just $36\%$. The
changes are even smaller for definitions~II or~III.

\begin{figure}
\bc
\hspace{-1.cm}
\resizebox{9cm}{!}{\includegraphics{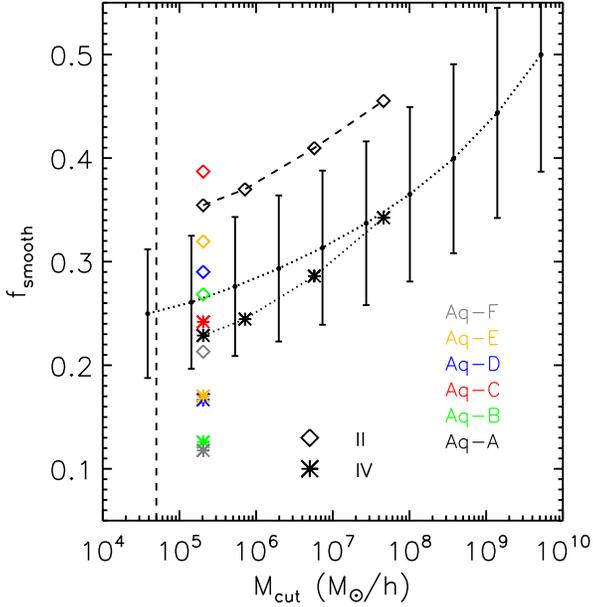}}\\%
\caption{The dependence of the diffuse accretion fraction,
  $f_{\rm smooth}$, on the mass resolution of the simulations, $M_{\rm
  cut}$, defined as the mass of a group of $20$ particles. Open
  diamonds and asterisks refer to diffuse accretion definitions II
  and IV, as described in the text and in the caption to
  Fig.~\ref{fig:diffacc}. Connected symbols refer to resolution levels
  2 through 5 for halo Aq-A. Colors indicate different halos, as
  labelled in the panel. Dots with error bars indicate the mean and
  {\it rms} scatter in several hundred Monte Carlo assembly histories
  constructed from excursion-set theory. The vertical dotted line
  indicates the value of $M_{\rm cut}$ for Aq-A-1, the highest
  resolution simulation in the Aquarius series.}
\label{fig:diffmcut}
\ec
\end{figure}

Finally, we consider the dependence of $f_{\rm smooth}$ on the mass
resolution of the simulations. For this, we use four different
realizations of halo Aq-A, from level 2 to level 5. The results are
shown in Fig.~\ref{fig:diffmcut} (criteria~II and~IV: open diamonds
and asterisks, respectively). As expected, there is a systematic
decrease in $f_{\rm smooth}$ with increasing resolution, measured in
Fig.~\ref{fig:diffmcut} by $M_{\rm cut}$, the mass of a group of $20$
particles.

Given that $f_{\rm smooth}$ depends on resolution, we need to ask how
secure is our estimate of this quantity from the simulations. We can
answer this by analyzing Monte Carlo merger trees built using the
excursion-set formalism, constrained at $z=0$ to make a halo of mass
comparable to those in the Aquarius set. In particular, we use the
algorithms of \citet{parkinson08} and
\citet{cole08}, which were tuned to match the N-body merger trees of
the Millennium Simulation
\citep{Springel2005}. This approach has the advantage that a cutoff
mass can be easily introduced in order to mimic the limited resolution
of a simulation \citep[see, e.g.,][]{Angulo2010}.

The excursion-set results are shown by the dotted line in 
Fig.~\ref{fig:diffacc} and by the connected dots in 
Fig.~\ref{fig:diffmcut}. The Monte Carlo trees,
when trimmed to match the resolution of the N-body simulations, give
results in good agreement with the simulations. The theoretical
calculation also confirms the large scatter in $f_{\rm smooth}$ seen
in the simulations. (The ``error bars'' on the Monte Carlo results
denote the {\it rms} scatter among several hundred realizations.)

The trends shown in Fig.~\ref{fig:diffmcut} imply that further
improvements in resolution would result in only small reductions in
the value of $f_{\rm smooth}$. Indeed, $f_{\rm smooth}$ seems to
depend more strongly on the particular definition adopted for smooth
accretion than on numerical resolution, at least for the $100$-million
particle halos we consider here. The vertical line in
Fig.~\ref{fig:diffmcut} shows the value of $M_{\rm cut}$ corresponding
to Aq-A-1, the best-resolved, billion-particle halo in the Aquarius
set. The Monte Carlo tree results suggest that its additional
resolution would result in only a very small decrease in the smoothly
accreted fraction.\footnote{We have not attempted the analysis
presented here for Aq-A-1 because of the formidable computational task
involved in building full merger trees for this simulation.}.

We conclude from this exercise that the substantial fraction of mass
found to be accreted smoothly in our simulations is a robust
result. We can therefore confidently rule out the claim by 
\citet{Madau2008} that at most $3\%$ of the mass of a galaxy-sized
halo can be supplied by smooth accretion. It is not clear at this
point what the cause of the disagreement is, but it is likely to be
related to the way in which these authors compute diffuse accretion
rather than to differences in the simulations themselves. For example,
the mass they regard as having been accreted in ``identifiable
subunits'' is just the sum of the masses of all progenitor halos that
contain particles that make it into the final system. A substantial
fraction of that summed mass includes particles that are {\it not}
part of the halo at $z=0$; this could have led
\citet{Madau2008} to overestimate the mass contributed by discrete
identifiable subunits and, therefore, to underestimate $f_{\rm
smooth}$.

Our conclusions are in agreement with those of
\citet{Angulo2010} (see also \citealt{Genel2010}): mergers {\it and}
smooth accretion are both defining features of the hierarchical
buildup of a CDM halo.

At redshifts $z\leq 6$ the intergalactic medium is fully photoionized
and gas is unable to collect in halos with maximum circular
velocity below about 15 km/s, corresponding to masses below roughly
$10^8M_\odot$ \citep{Okamoto08}.  This is well above the resolution limit of our
simulations, so the gas associated with these low-mass halos should be
considered to be diffusely accreted along with that associated with
the ``diffuse'' dark matter. Taking the limit at exactly
$10^8h^{-1}M_\odot$ for simplicity \footnote{The following numbers are
insensitive to this choice.}, we find that, on average, our halos
accrete 56\% of their baryons diffusely, with the numbers for
individual halos ranging from 46\% (Aq-F) to 64\% (Aq-C). The bulk of
baryonic accretion is thus predicted to be smooth rather than clumpy
for objects of Milky Way scale.

\begin{figure}
\bc
\hspace{-0.2cm}
\resizebox{8cm}{!}{\includegraphics{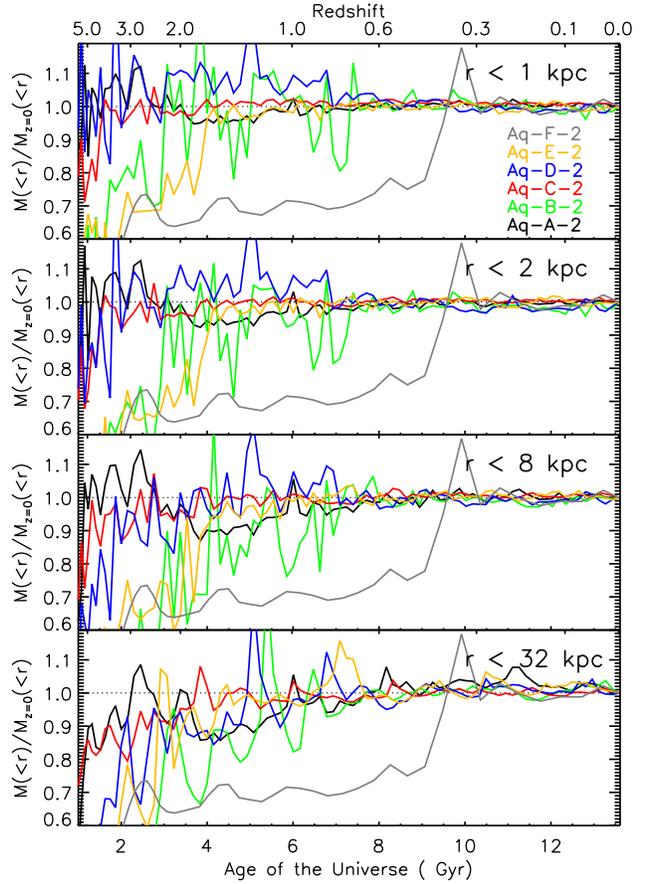}}\\%
\caption{Total mass enclosed within different physical radii, $ M (
  < r)$, for $ r= 1$; $2$; $8$; and $32$ kpc, as a function of cosmic
  time. Different colours correspond to different halos, as labelled in
  the figure. Masses in each panel are normalized to their values at
  $z=0$. Time is labelled at the bottom, redshift at the top. Except
  for halo Aq-F, which is the remnant of a recent major merger, the
  inner mass profile of Aquarius halos has been very stable for the
  past $5$-$6$ Gyr, a period comparable to the age of the Solar
  System.}
\label{fig:mrvst} \ec 
\end{figure}


\subsection{Evolution of the inner mass profile}
\label{SecEvInnProf} 

As Fig.~\ref{fig:mrgdiffacc} indicates, major mergers contribute, on
average, just under half of the particles in the inner $10 \hkpc$ of
Galactic halos. This is the region occupied by the luminous component
of the central galaxy, and it is thus interesting to analyze in detail
how the mass profile in this region evolves with time. A thin stellar
disk, for example, could react to clumpy addition of material by
thickening and becoming dyamically hotter, potentially violating
observations of the thin disk in the solar neighbourhood (see
e.g. \citealt{Benson04}).

We investigate the stability of the inner halo explicitly in
Fig.~\ref{fig:mrvst}, where we plot the mass enclosed within $1$, $2$,
$8$, and $32$ (physical) kpc\footnote{We assume $h=0.73$ in order to
compare with observations of the Milky Way.} from the centre of each
halo as a function of time. Each curve is normalized to the enclosed
mass at $z=0$. Except for halo F, which undergoes a major merger at
$z\sim 0.6$, all the halos show exceptionally stable inner mass
profiles over at least the past 5 Gyr ($z<0.6$), the age of the Solar
System. Five out of six Aquarius halos could, in principle, host a
disk as thin and cold as that of the Milky Way. Late-accreted mass
typically settles in the outskirts of a halo, thus allowing the
hierarchical growth of halos to be reconciled with the ubiquitous
presence of thin stellar disks.


\section{Summary and Conclusions}
\label{SecConcl}

We have analyzed the build-up of six $\Lambda$CDM halos simulated as part of
the Aquarius Project to study the influence of assembly history on
halo structure. We focus on the distributions of progenitor mass and
accretion time for particles at different radii at the final time, and
discuss various plausible definitions of accretion time, together with
the difficulties involved in estimating the total mass fraction
accreted smoothly. We compare simulations of the same halo carried out
with different resolution in order to assess the sensitivity of our
results to numerical limitations.

Although there is considerable variation from halo to halo, our
simulations exhibit a number of very clear trends. Our main
conclusions may be summarized as follows.
\begin{itemize}

\item There is a strong radial gradient in accretion time, which
  confirms that halos are built from the inside out.  Later accreting
  material settles farther from the centre of the halo; particles that
  today reside inside $10\hkpc$ are typically accreted $\sim 3$ Gyr
  earlier than particles that reside at $100\hkpc$ from the centre.

\item Similarly strong correlations exist between distance of a
  particle from halo centre and the mass at accretion of the
  progenitor halo which contained it. The innermost regions are
  dominated by particles brought in by massive clumps, ``major
  mergers'' with mass ratios exceeding $1$:$10$, as well as by those
  that joined the main progenitor at very early times ($z>6$).  Mass
  accreted diffusely and in minor mergers predominantly populates the
  more distant parts of the halo and dominates the total mass.

\item Minor mergers and diffuse accretion contribute approximately
equally to the mass of each halo at each radius, at least at the resolution of
our six Aquarius halos where minor mergers can be distinguished up to mass
ratios exceeding $10^6$:$1$. 

\item The inner mass profile of a halo is very stable at late times in
  systems that stay clear of major mergers. In five of the six
  Aquarius halos the profile within $32$ kpc barely changes in the
  past $5$-$6$ Gyr.

\item Diffuse accretion contributes a substantive fraction of the
  final mass of the halo, roughly $30$-$40\%$ in our simulations. This
  is a robust result compatible with expectations from excursion-set
  modeling.

\item Our analysis shows that some of the material accreted smoothly
  had previously been part of other collapsed structures, from which it was
  probably ejected by mergers. The same mechanism leads a
  fair fraction of particles in the main halo to cycle in and out of
  its main progenitor; at $z=0$ the cumulative mass of all particles
  ``associated'' in the past with the main progenitor exceeds the final
  mass of the halo by at least $20\%$.

\item After reionization, gas is unable to collect in dark halos with
  masses lower than about $10^8 \, h^{-1}\, M_\odot$.  As a result more than
  half of all the baryons associated with halos of Milky Way scale are
  expected to be accreted smoothly, rather than in clumps.
\end{itemize}

These results emphasize the dynamic nature of halo buildup and provide
insight into the radial structure of a halo and the history of its
assembly process. The view that emerges highlights some misconceptions
regarding hierarchical growth. CDM halos are not passive repositories
where mass is added in continuous but discrete events, but rather
lively systems that can lose as well as gain material throughout their
lifetimes. Diffuse accretion, recurring infall, escape and fallback,
are all processes that play important roles in the build-up of CDM
halos.

\section*{Acknowledgements}

The simulations of the Aquarius Project were carried out at the
Leibniz Computing Center, Garching, Germany, at the Computing Centre
of the Max-Planck-Society in Garching, at the Institute for
Computational Cosmology in Durham, and on the `STELLA' supercomputer
of the LOFAR experiment at the University of Groningen. We thank Shaun
Cole for providing us the code to produce Monte Carlo merger trees, and  
Mike Boylan-Kolchin for useful comments and careful reading of the 
manuscript. JW acknowledges a Royal Society Newton International 
Fellowship, CSF a Royal Society Wolfson Research Merit Award and AH 
support from a VIDI grant by Netherlands Organisation for Scientific 
Research (NWO). AH acknowledges funding from the European Research 
Council under ERC-StG GALACTICA-240271. This work was supported by an 
STFC rolling grant to the Institute for Computational Cosmology.

\bsp
\label{lastpage}

\bibliographystyle{mn2e}
\bibliography{Aqu_mah}

\end{document}